\documentclass[preprint,aps]{revtex4}
\usepackage{graphicx}
\usepackage{booktabs}
\usepackage[usenames, dvipsnames]{color}
\usepackage{multirow}
\usepackage{setspace}
\usepackage{amsmath, amsthm, amssymb}
\usepackage{rotating}
\usepackage[ pdftex, plainpages = false, pdfpagelabels, 
                 pdfpagelayout = useoutlines,
                 bookmarks,
                 bookmarksopen = true,
                 bookmarksnumbered = true,
                 breaklinks = true,
                 linktocpage=all,
                 pagebackref=false,
                 colorlinks = true,
                 linkcolor = BrickRed,
                 urlcolor  = blue,
                 citecolor = BrickRed,
                 anchorcolor = green,
                 hyperindex = true,
                 hyperfigures
                 ]{hyperref} 
                 
\newcommand{\comment}[1]{}

\newcounter{figref}

\begin{document}

\title{\href{http://necsi.edu/research/social/foodprices/updatejuly2012/}{UPDATE July 2012 --- The Food Crises: The US Drought}} 
\date{July 23, 2012}  
\author{Marco Lagi, Yavni Bar-Yam and \href{http://necsi.edu/faculty/bar-yam.html}{Yaneer Bar-Yam}}
\affiliation{\href{http://www.necsi.edu}{New England Complex Systems Institute} \\ 
238 Main St. Suite 319 Cambridge MA 02142, USA \vspace{2ex}}

\begin{abstract}
Recent droughts in the midwestern United States threaten to cause global catastrophe driven by a speculator amplified food price bubble. Here we show the effect of speculators on food prices using a validated quantitative model that accurately describes historical food prices. During the last six years, high and fluctuating food prices have lead to widespread hunger and social unrest. While a relative dip in food prices occurred during the spring of 2012, a massive drought in the American Midwest in June and July threatens to trigger another crisis. In a previous paper, we constructed a model that quantitatively agreed with food prices and demonstrated that, while the behavior could not be explained by supply and demand economics, it could be parsimoniously and accurately described by a model which included both the conversion of corn into ethanol and speculator trend following. An update to the original paper in February 2012 demonstrated that the model previously published was predictive of the ongoing price dynamics, and anticipated a new food crisis by the end of 2012 if adequate policy actions were not implemented. Here we provide a second update, evaluating the effects of the current drought on global food prices. We find that the drought may trigger the expected third food price bubble to occur sooner, before new limits to speculation are scheduled to take effect. Reducing the amount of corn that is being converted to ethanol may address the immediate crisis. Over the longer term, market stabilization requires limiting financial speculation. 
\end{abstract}

\maketitle

The current global crisis in food prices and the vulnerability of the limited food supply to environmental and other disruptions is a matter of ongoing concern \cite{worldbank_foodcrisis}. Recent food price peaks in 2007-08 and 2010-11 have resulted in food riots and are implicated in triggering widespread revolutions known as the Arab Spring \cite{food_crises}. The underlying vulnerability of the global food supply system is being tested again this summer by a severe drought in the Midwestern United States which is responsible for a large portion of the global food supply \cite{drought_monitor,NOAA_drought_June,Reuters_drought,source_grains}.

In a paper published in September 2011 \cite{food_prices}, we built a quantitative model that, for the first time, was able to precisely match the monthly FAO food price index over the last 8 years. The model showed that, of all the factors proposed to be responsible for the recent dramatic spikes and fluctuations in global food prices, 
rapid increases in the amount of 
corn-to-ethanol conversion and speculation on futures markets were the only factors which could justifiably be held responsible. 
Each of these causes in turn results from particular acts of government intervention or deregulation. 
Thus, while the food supply and prices may be vulnerable to global population increases and environmental change, the existing price increases are due to specific governmental policies. 
In order to prevent further crises in the food market, we recommended the halting of government support for ethanol conversion and the reversal of commodities market deregulation, which enables unlimited financial speculation.

Since the publication of our analysis, a few changes in these directions have been made.
At the end of 2011, ethanol subsidies were allowed to expire \cite{Pear2012,Krauss2011}. However, a government-guaranteed demand for 37\% of the US corn crop is still in place \cite{Smith2012}. It is unclear what effect this partial change in policy will have on the percentage of corn converted to ethanol, which is currently about 40\%. 
New position limits on speculative activity by the US Commodities Futures Trading Commission are scheduled to come into effect by the end of 2012, as required by the Dodd-Frank Act \cite{doddfrank,FedReg}. 
It remains to be seen how effective these new regulations will be, as there are those who consider them too watered-down \cite{PattersonTrindle2011}, and market participants are seeking to dilute them further \cite{PattersonTrindle2011,Trindle2012,Brush2012,Donahue2010,Damgard2011}.

In a subsequent update published in February 2012 \cite{feb_update}, we showed that the model continued to fit current data, up to January 2012, which had not been available at the time of the construction of the  
model. We further observed that extrapolating model prices into the future yielded the prediction of another speculative bubble starting by the end of 2012 and causing food prices to rise even higher than recent peaks.

This season, the American Midwestern agricultural region has experienced debilitating droughts and high temperatures, the most severe in at least 50 years, leading to rapidly rising corn and wheat prices in anticipation of a poor yield \cite{NOAA_drought_June,Reuters_drought}. 
Here we include this as a shock in our established model. We find that through the mechanism of speculative activity, the drought may trigger a third massive price spike to occur earlier than otherwise expected, beginning immediately, and sooner than could be prevented by the anticipated new regulations. This spike 
may raise prices well beyond an increase
justified by the reduced supply caused by the droughts.

\begin{figure}[t]
\refstepcounter{figref}\label{fig:food_pert}
\href{http://necsi.edu/research/social/foodprices/updatejuly2012/food_7_2012.pdf}{\includegraphics[width=1\linewidth]{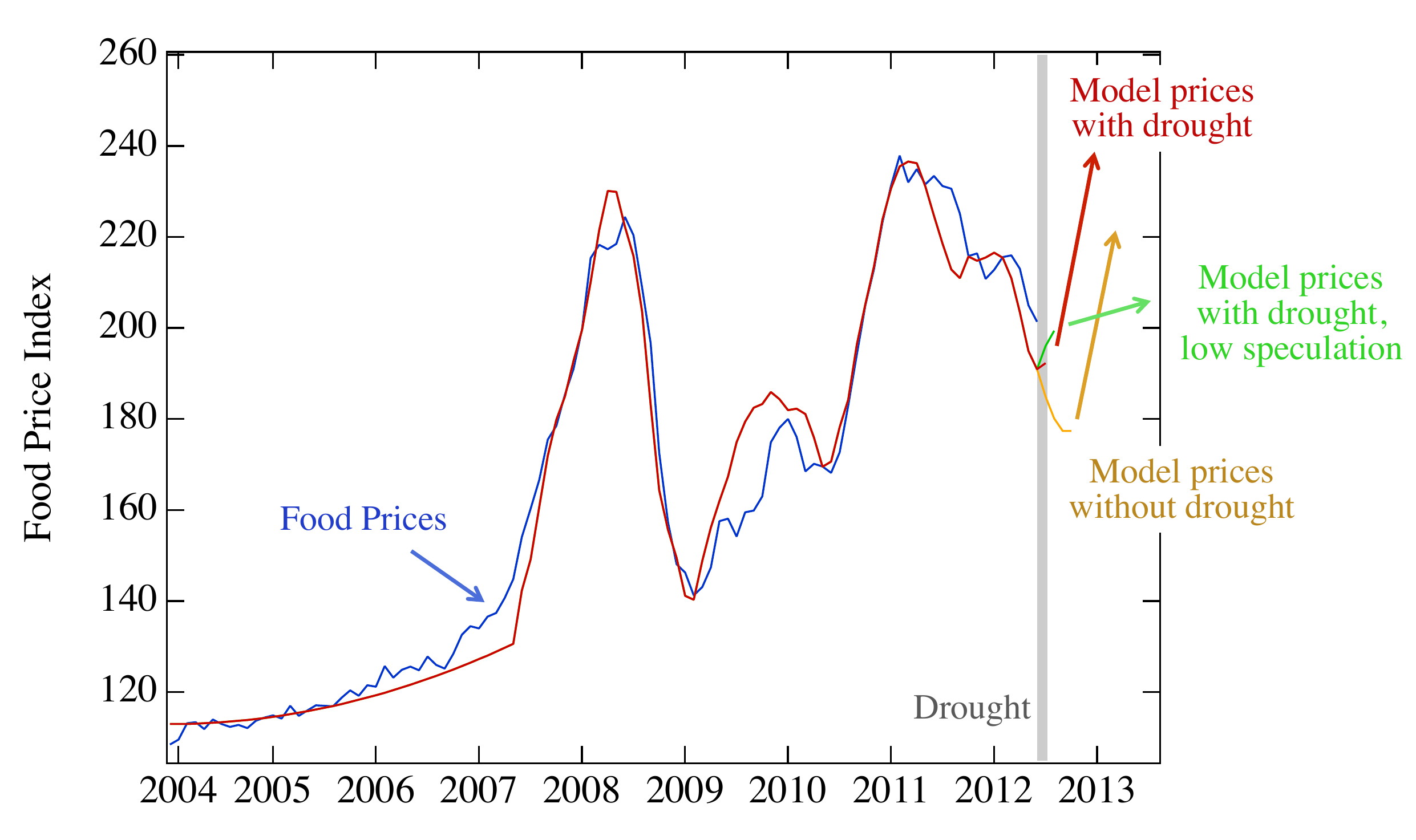}}
\caption{\textbf{Food prices and model simulations} - The FAO Food Price Index (blue solid line) \cite{source_fao}, and the speculator and ethanol model \cite{food_prices}, without the effects of the current drought (yellow line) with the effects of the current drought (red line), and with the effects of the drought, but with speculation reduced (green line). In all cases, corn-to-ethanol conversion is considered to be constant after Jan 2012 and stock prices and bond prices are considered to be constant after July 2012. For the red and green line, drought is modeled as a shock in equilibrium prices (+3\%) in July 2012. In all cases, the new optimized parameters for the fit up to July 2012 are: $k_{sd} = 0.089$, $k_{sp} = 1.25$, $\mu_{equity}\gamma_0 = -0.074$, $\mu_{bonds}\gamma_0 = -15.4$. The speculation parameter after July 2012 is reduced to $k_{sp} = 0.3$ for the green line.}
\end{figure}
 
In Fig. \ref{fig:food_pert} we plot our model predictions for different scenarios. 
The central comparison is between the drought triggered speculative bubble (red line) compared to the same shock with reduced speculation (green line). While the drought causes only a limited price shock, the impact on prices is amplified by the speculative activity. Fig. \ref{fig:food_pert} shows the quantitative impact of speculators given the current level of market speculation as validated by prior analysis of food prices.

\begin{figure}[t]
\refstepcounter{figref}\label{fig:grains}
\href{http://necsi.edu/research/social/foodprices/updatejuly2012/food_7_2012_2.png}{\includegraphics[width=0.8\linewidth]{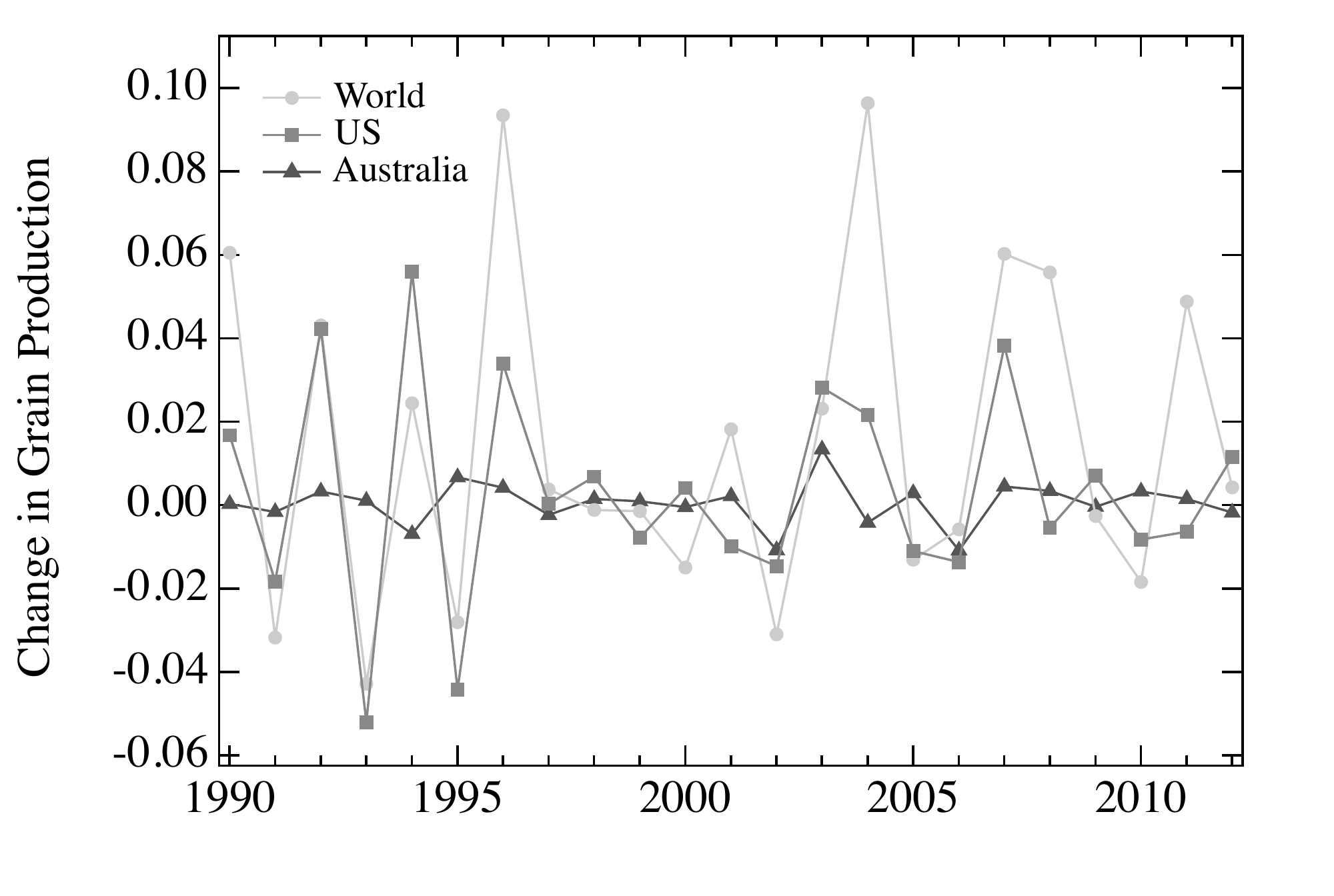}}
\caption{\textbf{Correlations of changes in grain production} - Comparison of change in world (circles), US (squares) and Australian (triangles) grain production as a fraction of total world production by weight \cite{source_grains}. The correlation is small between world and Australia production changes (0.17), but high between world and US production changes (0.71).}
\end{figure}

The speculative bubble is modeled starting from a price shock driven by the drought, which is expected to occur based upon existing grain price increases \cite{Reuters_drought} though it has not yet been specified by the FAO. The price increase then causes the upcoming price spike to come sooner than would have otherwise occurred. The level of earlier riot-inducing bubbles is reached before the end of 2012 and prices continue to rise much higher. Without the drought (yellow line), the rise in prices would be just as dramatic, but is predicted to occur several months later, possibly in time for the new regulations to prevent it. On the other hand, if speculation were to be curbed immediately, starting from July 2012, the model shows 
(green curve) that the price increase due to the drought would be far smaller, and would not lead to another dramatic price spike.
An alternative intervention, eliminating the government mandated ethanol quota for this year \cite{worstall2012}, would would result in a new market shock and could cause a sudden drop in prices. This may alleviate the immediate concerns though its effect is subject to speculator-driven bandwagon effects.

We note that in our original paper \cite{food_prices}, we evaluated the suggestion 
that droughts in Australia had been responsible for the increase of food prices in 2008. We concluded that they could not have been, because the Australian production of grain is not a sufficiently large portion of the global production of grain. The same analysis yields a different answer for the US drought. US production of grain is consistently an order of magnitude larger than that of Australia \cite{source_grains}. Moreover, changes in the global production of grain 
correlate with changes in US production (see Fig. \ref{fig:grains}). The two time series have a Pearson's correlation coefficient over the last 22 years of $\rho = 0.71$ compared to $\rho = 0.17$ for the correlation with Australia's time series.

We thank Peter Timmer for helpful conversations. This work was supported in part by AFOSR under grant FA9550-09-1-0324, ONR under grant N000140910516.

\end{document}